\documentclass[aps,prb,twocolumn,superscriptaddress,floatfix,longbibliography]{revtex4-2}

\usepackage{amsmath,amssymb} 
\usepackage{bm} 
\usepackage{graphicx} 
\usepackage{appendix}
\usepackage{enumitem} 
\setlist{noitemsep,leftmargin=*,topsep=0pt,parsep=0pt}
\usepackage[english]{babel}
\usepackage{xcolor} 
\definecolor{lightgray}{gray}{0.6}
\definecolor{medgray}{gray}{0.4}
\usepackage{natbib}
\newif\ifptitle
\newif\ifpnumber
\newcounter{para}

\ptitletrue  
\pnumbertrue  

\begin{document}

\title{Site-selective observation of spin dynamics of a Tomonaga-Luttinger liquid in frustrated Heisenberg chains}

\author{Diep Minh Nguyen}
\affiliation{Department of Physics, Nagoya University, Nagoya 464-8601, Japan.}
\author{Azimjon A. Temurjonov}
\affiliation{Department of Physics, Nagoya University, Nagoya 464-8601, Japan.}
\author{Daigorou Hirai}
\affiliation{Department of Applied Physics, Nagoya University, Nagoya 464-8603, Japan.}
\author{Zenji Hiroi}
\affiliation{Institute for Solid State Physics, University of Tokyo, Kashiwa, 277-8581, Japan.}
\author{Oleg Janson}
\affiliation{Institute for Theoretical Solid State Physics, Dresden, 01069, Germany.}
\author{Hiroshi Yasuoka}
\affiliation{Max Plank Institute for Chemical Physics of Solids, 01187 Dresden, Germany.}
\author{Taku Matsushita}
\affiliation{Department of Physics, Nagoya University, Nagoya 464-8601, Japan.}
\author{Yoshiaki Kobayashi}
\affiliation{Department of Physics, Nagoya University, Nagoya 464-8601, Japan.}
\author{Yasuhiro Shimizu}
\altaffiliation[Present address: ]{Department of Physics, Shizuoka University, Shizuoka, 422-8529, Japan}
\affiliation{Department of Physics, Nagoya University, Nagoya 464-8601, Japan.}

\date{\today}

\begin{abstract}
Low-energy spin dynamics is investigated by $^{35}$ Cl NMR measurements in a frustrated antiferromagnet Ca$_3$ReO$_5$Cl$_2$. The local spin susceptibility measured with the Knight shift behaves as a one-dimensional Heisenberg antiferromagnet and remains constant down to low temperatures, as expected in a gapless Tomonaga-Luttinger liquid. The nuclear spin-lattice relaxation rate $T_1^{-1}$ demonstrates a slowing down of atomic motions and a power-law evolution of spin correlation. The Luttinger parameter is enhanced in a site-selective manner depending on the form factor of dynamical spin susceptibility. The strong anisotropy of $T_1^{-1}$ reflects the strong spin-orbit coupling through Dzyaloshinskii-Moriya interaction. The ground state exhibits an incommensurate antiferromagnetic ordering with low-lying magnon excitations. 
\end{abstract}

\maketitle

\section{\label{sec:Start}Introduction}
A Tomonaga-Luttinger liquid (TLL) is characterized by fractional gapless excitations and bosonic collective modes in one-dimensional (1D) localized spin chains \cite{Haldane1981,Giamarchi,Sachdev2011}. The spin correlation function of TLL follows a universal power law. One of the observables is the nuclear spin-lattice relaxation rate $T_1^{-1}$, which measures the dynamical spin susceptibility summed over the wave vector space in a low-energy limit. It approximately obeys the power-law temperature $T$ dependence  \cite{Chitra1997,Bocquet2001,Hikihara,Sato2009,Sato2011,Coira2016}: 
\begin{equation}
   T_1^{-1} \propto T^{1/2K_{\sigma}-1}, 
\end{equation}
where $K_{\sigma}$ is one of the TLL parameters. A spin-1/2 Heisenberg chain yields $K_\sigma = 1/2$, in which $T_1^{-1}$ becomes $T$-invariant \cite{Sachdev1994,Takigawa1991, Takigawa1996,Thurber2001}. $K_{\sigma}$ depends on the Ising anisotropy and the magnetic field strength respectively acting as the interaction and chemical potential of spinless fermions \cite{Giamarchi}. As three-dimensional (3D) coupling between the chains sets in, the system heads toward long-range magnetic ordering with the effectively enhanced $K_\sigma$ above the transition temperature $T_{\rm N}$ \cite{Dupont2018, Horvatic2020}. On the spin-1/2 ladder and spin-1 chain exhibiting a nonmagnetic ground state, the spin correlation measured with $K_\sigma$ sensitively depends on the magnetic field \cite{Giamarchi1999,Thurber2000, Klanjsek2008, Bouillot2011, Mukhopadhyay2012, Jeong2016, Moller2017, Dupont2016}. 

The anisotropic triangular antiferromagnet connects 1D TLL to two-dimensional (2D) quantum spin liquid as a function of the ratio of the interchain interaction $J^\prime$ and the intrachain interaction $J$ \cite{Giamarchi, Sachdev_Senthil1994,Yunoki2006, Heidarian2009, Hauke2013}. Geometrical frustration reduces the interchain correlation and hence the system dimensionality \cite{Weng, Hayashi2007}. The TLL phase is stabilized up to the exchange anisotropy $J^\prime/J \approx 0.3$ for the interchain exchange coupling $J^\prime$ and the intrachain coupling $J$ \cite{Weng, Yunoki2006, Starykh2007, Hayashi2007,Heidarian2009, Hauke2013}. A spiral magnetic order occurs for $J^\prime/J > 0.3$-$0.7$ on the triangular lattice. The quantum spin liquid, if any, near the isotropic point $J^\prime/J \approx 1$ may involve a tiny spin gap and topological order \cite{Sachdev1994,Wen}, which is distinct from the renormalized 1D liquid phase. 

\begin{figure}
\centering
\includegraphics[width=8cm]{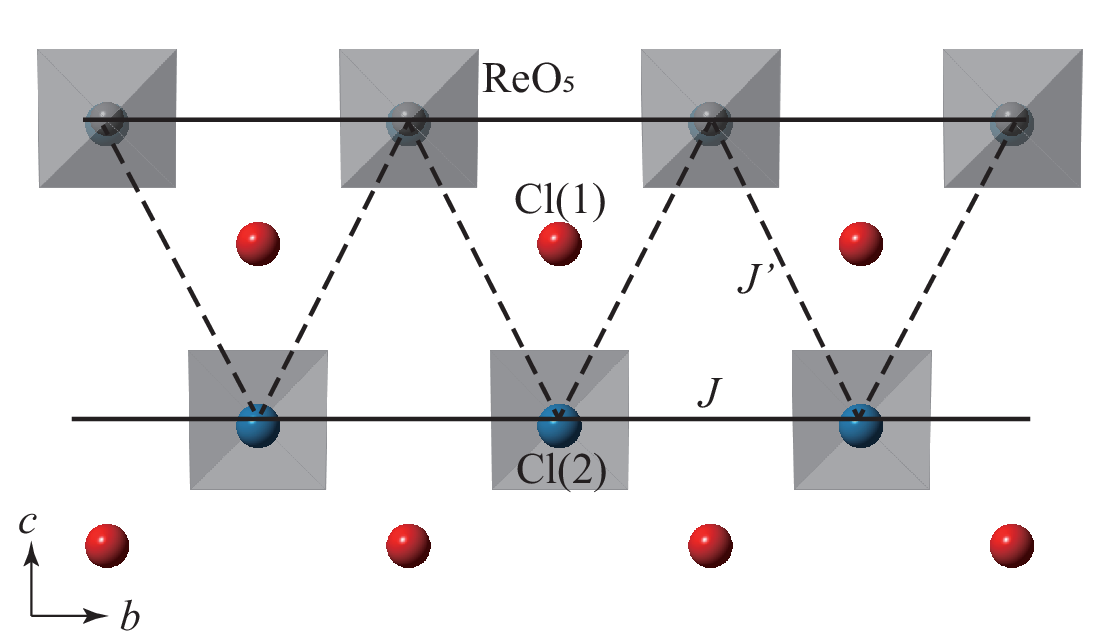}
\hfill
\caption{Crystal structure of Ca$_{3}$ReO$_{5}$Cl$_{2}$ ($Pnma$) without Ca atoms in a half of the unit cell along the $a$ axis for simplicity. The intrachain interaction $J$ and the interchain one $J^{\prime}$ are shown by the solid and dash lines, respectively. Cl(1) and Cl(2) sites are respectively located between the ReO$_5$ chains and on the ReO$_5$ pyramid, as viewed from the $a$ axis. The symmetry operation of Re ($4c$) and two Cl sites ($4c$) are expressed in combinations of the $a$-glide normal to the $c$ axis and the twofold screw along the three axes, where the fractional atomic coordinates are Re = (0.6855, 3/4, 0.0824), Cl(1) = (0.4639, 1/4, 0.2102), Cl(2) = (0.4854, 3/4, 0.4089) \cite{Hirai2017}. }
\label{fig:crystal}
\end{figure}

An example of the spin-1/2 anisotropic triangular antiferromagnets, Cs$_2$CuCl$_4$ ($J^\prime/J = 0.3$), exhibits magnetic orders and TLL behavior with bound spinons, depending on the magnetic field strength and direction, as observed by NMR and inelastic neutron-scattering measurements \cite{Coldea2001, Coldea2003, Kohno2007, Starykh2007, Starykh2010, Vachon2006,Vachon2011}. A $5d$ transition-metal oxychloride Ca$_3$ReO$_5$Cl$_2$ (CROC) is served as another example of the spin-1/2 quasi-1D antiferromagnet with the anisotropic triangular lattice, as shown in Fig. 1 \cite{Hirai2017, Hirai2018,Hirai2024}. CROC forms an orthorhombic $Pnma$ lattice, where the ReO$_5$ square-pyramid with the $5d_{xy}$ orbital forms a chain along the $b$ axis. Despite the strong spin-orbit coupling of Re ions, the orbital moment is quenched under the asymmetric ligand field and perturbatively contributes to the ground state through the Dzyaloshinskii-Moriya (DM) interaction ${\bf D}\cdot{\bf S}_i \times {\bf S}_{i+1}$, where ${\bf D}$ is parallel to the $c$ axis in the $ac$ mirror plane. The interchain coupling within the $bc$ plane constructs an anisotropic triangular lattice, as shown in Fig. \ref{fig:crystal}. The anisotropy of the triangular lattice has been evaluated as $J^\prime$/$J$ = 0.25 from the high-field magnetization \cite{Zvyagin2022}, consistent with that of the calculation of density functional theory (DFT) ($J^\prime$/$J$ = 0.295) \cite{Hirai2018, Choi2021}. At low temperatures, $\chi$ behaves as a spin-1/2 1D antiferromagnetic Heisenberg model with $J/k_{\rm B}$ = 41 K \cite{Hirai2018,Zheng2005}, which implies a reduction of the dimensionality. The ground state eventually exhibits a long-range magnetic order below $T_{\rm N}$ = 1.13 K. Thus, one can investigate the spin correlation of the TLL state featured by $K_\sigma$ in an extensive temperature range of $T_{\rm N} < T < J$. Below $T_{\rm N}$, magnon excitation may coexist with bound spinons in the high-energy dispersion observed by the inelastic neutron scattering \cite{Nawa2020}. 

Here we investigate the spin excitation and structure on the quasi-1D CROC with site-selective Cl NMR spectroscopy. We determine the Knight shift and electric field gradient (EFG) tensors for two Cl sites and then compare the result with the calculation. $K$ and $T_1^{-1}$ measurements respectively uncover static and dynamic spin susceptibilities of CROC in the TLL regime at low temperatures. We show remarkable anisotropy and site dependence of $T_1^{-1}$, which are discussed in terms of the DM interaction and the wave-vector dependence of the form factor of the anisotropic dynamical spin susceptibility. 

\section{\label{sec:Exp}Method}

The single crystals of CROC were grown by a flux method with a mixture of CaO, ReO$_3$, and CaCl$_2$ in a quartz ampule \cite{Hirai2017}. The typical dimensions of the crystal were 0.5 $\times$ 5 $\times$ 2 mm$^3$. We performed $^{35}$Cl and $^{37}$Cl (nuclear spins $I = 3/2$) NMR measurements on a single crystal of CROC under the steady magnetic field $H_0$. We utilized a dual-axis rotator for the angular dependence measurement above 1.5 K and a single-axis one in a $^3$He cryostat below 1.5 K. Frequency-swept NMR spectra were obtained from spin-echo signals taken by a 0.3 MHz step using a pulse sequence $\pi/2-\tau-\pi$ with $\pi/2$ = 1.5--2.0 $\mu s$ and $\tau$ = 50--100 $\mu$s. The Knight shift $K$ was evaluated from the central resonance frequency by subtracting the higher-order quadrupole contribution with the exact diagonalization of the nuclear-spin Hamiltonian in Appendix \ref{quadrupole}. The spin-lattice relaxation rate $T_1^{-1}$ was measured with a saturation recovery method, in which nuclear magnetization $M(t)$ obeys
 $ M(t) = M_0 - M_0 \left[0.1e^{-t/T_1} + 0.9e^{-6t/T_1}\right].$ 
The spin-echo decay rate $T_2^{-1}$ was obtained from $M(2\tau) \propto {\rm exp}(-2\tau/T_2)$ for two Cl sites. 

The nuclear quadrupole frequency was calculated with \textsc{FPLO} version 22 \cite{Koepernik1999} for the standard basis (S), the extended basis (D), and the extended basis with additional $4f$ (D4f) basis sets. The values presented in the manuscript were calculated using the D4f basis, which provides the highest accuracy, on a mesh of $8 \times 16 \times 8 k$ points. Relativistic effects were neglected
full-relativistic calculations on sparser meshes yielded nearly identical (differing by less than 4\%) quadrupole frequencies. Since there are many heavy atoms in the structure, we use the local density approximation (LDA) functional \cite{Perdew1992}. 

\section{Experimental results}
\subsection{Local spin susceptibility and magnetic order}

\begin{figure}
\centering
\includegraphics[width=8.5cm]{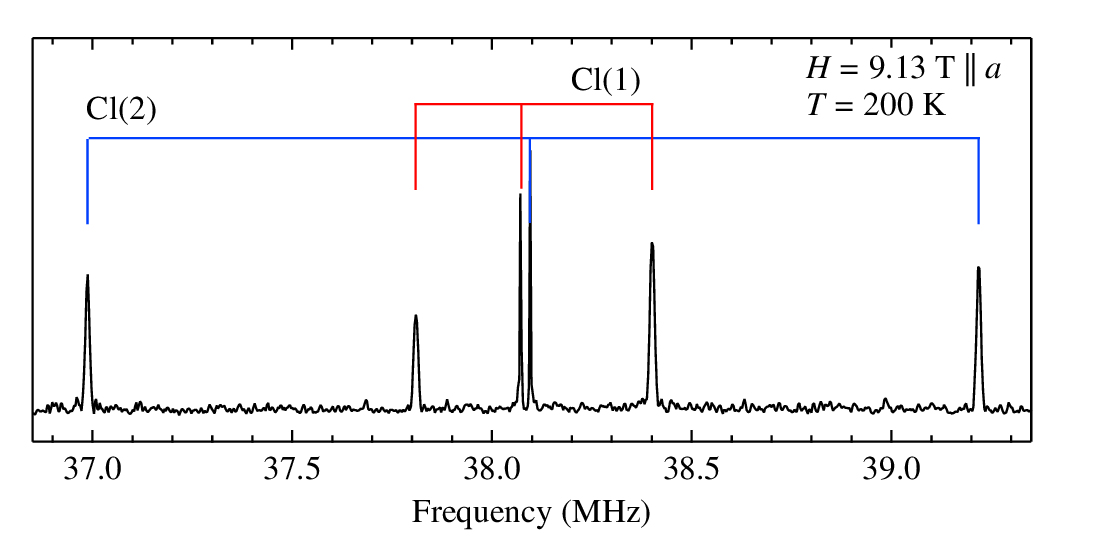}
\caption{$^{35}$Cl NMR spectrum of CROC at 200 K and $H_0$ = 9.13 T along the $a$ axis. The linewidth is $\approx$ 2 kHz for central lines. For the spectrum assignment we refer to the angular dependence of the nuclear quadrupole slitting frequency $\delta \nu$ and the central frequency shift in Appendix \ref{quadrupole}.}
\label{fig:spectrum}
\end{figure}

The $^{35}$Cl NMR spectrum of CROC consists of extremely sharp resonance lines coming from two Cl sites, Cl(1) and Cl(2), under magnetic field along the $a$ axis, as shown in Fig. \ref{fig:spectrum}. For $I= 3/2$, the spectrum from each Cl site splits into three due to the electric-quadrupole interaction between the nuclear quadrupole moment $Q$ and the EFG at the Cl site. The number of resonance lines doubles as the inversion symmetry is broken away from the $ab$ and $bc$ planes, which allows us the accurate field alignment along the crystal axes within $1^\circ$. The quadrupole splitting $\delta \nu$ and the central frequency are plotted in Fig. 8 of Appendix \ref{quadrupole}. The angular dependence is compared with the LDA calculation based on the crystal structure of CROC, leading to the unambiguous site assignment of the resonance lines into Cl(1) and Cl(2), as shown in Fig. \ref{fig:spectrum}. 

\begin{figure}
\includegraphics[scale=0.5]{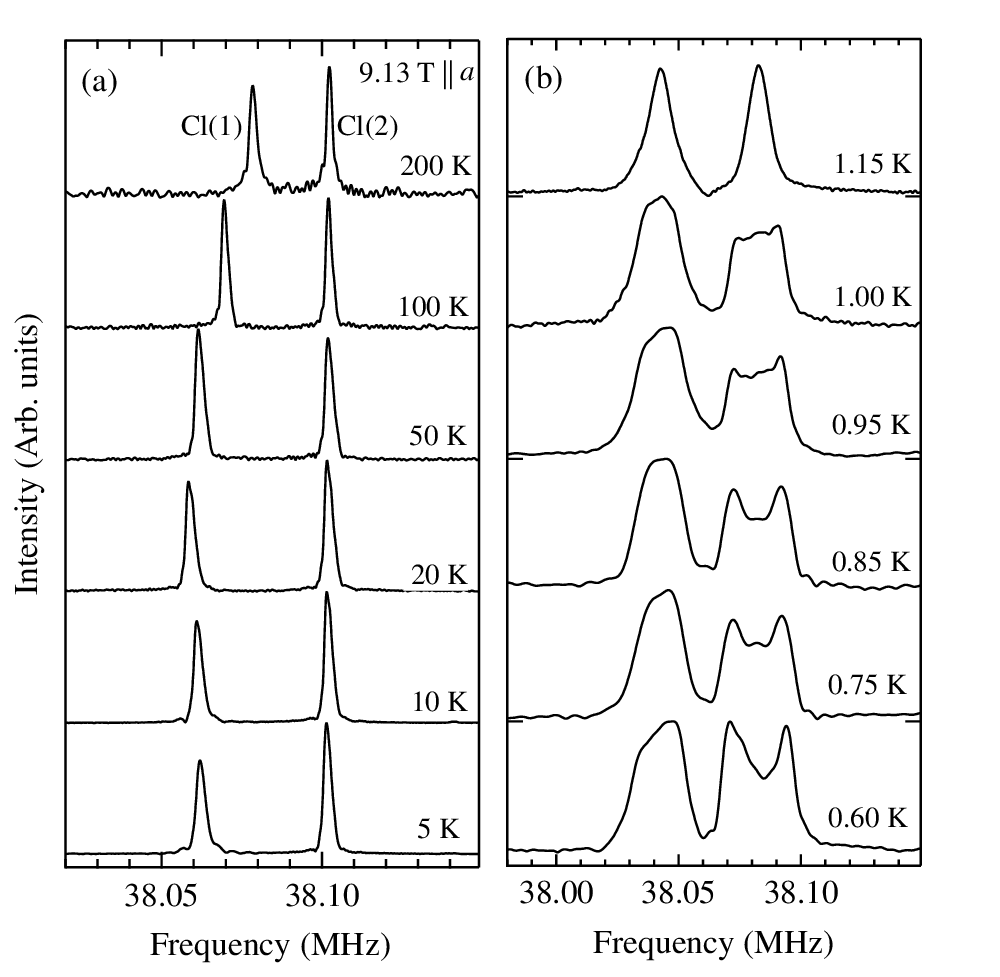}
\caption{Temperature dependence of the central $^{35}$Cl NMR spectrum for the single crystal of Ca$_3$ReO$_5$Cl$_2$ at 9.13 T along the $a$ axis (a) above and (b) below $T_{\rm N}$.} 
\label{fig:He3}
\end{figure}

The NMR spectrum displays a site-dependent shift upon cooling, as shown in Fig. \ref{fig:He3}(a). The low-frequency spectrum from Cl(1) exhibits a downward shift, while the higher one from Cl(2) stays at nearly the same position. Since $\chi$ exhibits Curie-Weiss paramagnetic behavior at high temperatures, the downward shift indicates the negative hyperfine coupling constant for Cl(1). Surprisingly, the linewidth remains sharp down to low temperatures, indicating a high-quality crystal free from the magnetic inhomogeneity arond free spins. 

\begin{figure}
\centering
\includegraphics[width=8.5cm]{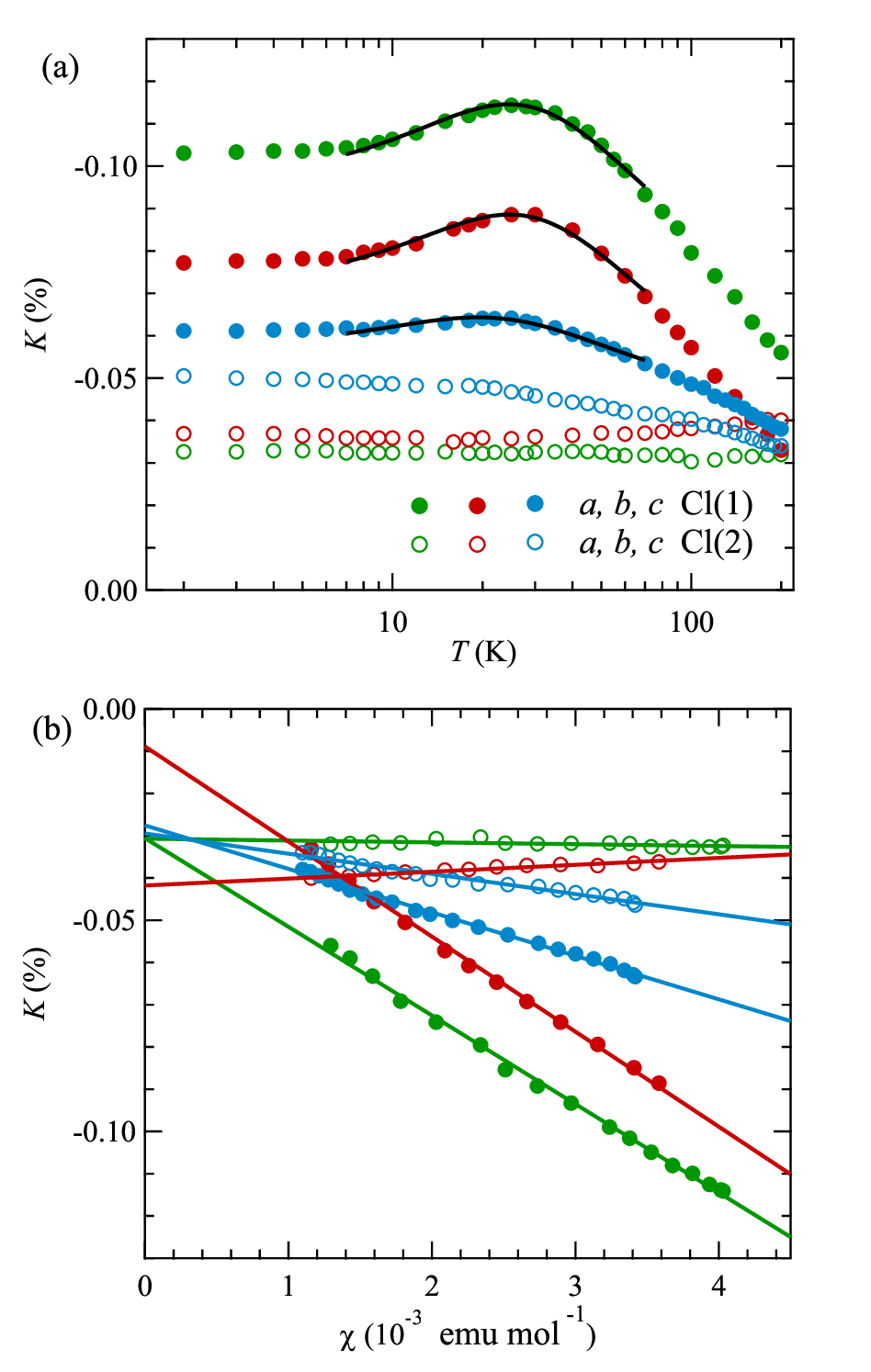}
\caption{(a) Knight shift $K$ measured as a function of temperature $T$ along the crystal axes for two Cl sites. Note that the vertical axis is taken upside down. Solid curves are the Bonner-Fisher fitting with $J$ = 38 K. (b) $K$ plotted against the bulk magnetic susceptibility $\chi$ as an implicit function of $T$. Solid lines are linear fitting results for $T > 28$ K. }
\label{fig:K-Chi}
\end{figure}

The Knight shift $K$ obtained from the central resonance line scales to the local spin susceptibility $\chi_{ii}$ along the crystal $i$ axis; $K_{ii} = \frac{A_{ii}}{N\mu_{\rm B}}\chi_{ii}$ ($i = a, b, c$), where $A_{ii}$ is the diagonal hyperfine coupling, $N$ the Avogadro number, and $\mu_{\rm B}$ the Bohr magneton. Figure \ref{fig:K-Chi} shows the $T$ dependence of $K$ for Cl(1) and Cl(2), denoted as $K(1)$ and $K(2)$, respectively. $-K(1)$ increases upon cooling, showing a broad peak around 28 K. It becomes nearly constant at low temperatures. The temperature dependence of $-K(1)$ for each crystal axis fits to the Bonner-Fisher model with $J/k_{\rm B}$ = 38 K ($7 < T < 70$ K) \cite{BonnerFisher} in agreement with the result of $\chi$ ($J/k_{\rm B} = 41.3$ K) \cite{Hirai2018}. In contrast with the bulk $\chi$, the local spin susceptibility obtained from $K(1)$ excludes an impurity contribution and extracts the intrinsic residual spin susceptibility down to low temperatures ($3.4 \times 10^{-3}$ emu/mol), which corroborates the gapless TLL state. The spin susceptibility of TLL is expressed as $\chi(0) = (g\mu_{\rm B})^2K_\sigma/(\pi v)$ using the spin velocity $v = J\pi/2$ at $T=0$ \cite{Eggert1994,Ninios2012}. Using the $T$-linear term of the specific heat, $\gamma = 115$ mJ/(K mol) \cite{Hirai2018}, the Wilson ratio $R_W = (\pi k_{\rm B}/g\mu_{\rm B})^2(4\chi/3\gamma)$ is obtained as 2.46. It is compared with $R_W = 4K_\sigma$ expected for TLL spin chains \cite{Ninios2012}. Using $K_\sigma =0.61-0.87$ for Cl(1) from $T_1^{-1}$ as described below (Table. \ref{table:K_sigma}), we can independently obtain $R_W = 2.44$--3.48, consistent with the Knight shift. 

Since $\chi$ is nearly isotropic \cite{Hirai2018}, the anisotropy of $K(1)$ comes from the anisotropic hyperfine coupling constant governed by transfer or dipole interactions. $K(1)$ linearly scales to $\chi$ above 28 K, as seen in the $K(1)-\chi$ plot [Fig. \ref{fig:K-Chi}(b)]. The components of the hyperfine coupling tensor are evaluated from the linearity as ($A_{aa}$, $A_{bb}$, $A_{cc}$, $A_{ac}$) = ($-0.12$, $-0.13$, $-0.058$, 0.13)T/$\mu_{\rm B}$ for Cl(1). Here, $A_{ac}$ is evaluated from the angular dependence of Knight shifts. The diagonalization yields the principal components of the hyperfine coupling ($A_{XX}$, $A_{YY}$, $A_{ZZ}$) = (0.04, $-0.13$, $-0.22$)T/$\mu_{\rm B}$ for Cl(1). The anisotropy is not explained by the dipolar interaction ($\sum_\alpha A_{\alpha\alpha} = 0$), and hence comes from the transferred hyperfine interaction. In contrast, $K(2)$ along the $a$ and $b$ axes very weakly depends on $T$, indicating the tiny hyperfine coupling constants. We obtained ($A_{aa}$, $A_{bb}$, $A_{cc}$, $A_{ac}$) = ($-0.002$, 0.01, $-0.03$, 0.14)T/$\mu_{\rm B}$ and ($A_{XX}$, $A_{YY}$, $A_{ZZ}$) = (0.12, 0.01, $-0.15$)T/$\mu_{\rm B}$ for Cl(2). The remarkable site dependence reflects the transferred hyperfine paths originating the anisotropic Re $5d_{xy}$ orbital, as discussed in Appendixes. 

Below $T_{\rm N}$ = 1.1 K, the NMR spectrum broadens and changes to a double horn shape, as shown in Fig. \ref{fig:He3}(b). It shows an emergence of the spontaneous local field below $T_{\rm N}$. One can exclude a possible collinear magnetic ordering that involves a discrete splitting of the NMR spectrum by the staggered hyperfine field. Despite a hyperfine coupling constant of Cl(1) greater than Cl(2), the splitting amplitude is smaller for Cl(1) along the $a$ axis. This means that the local fields at Cl sites are generated through the off-diagonal hyperfine coupling from the Re magnetic moments oriented perpendicular to the $a$ axis. The double horn shape is typically observed in incommensurate magnetic orders such as corn and spiral orders \cite{Starykh2010,Nawa2013}. In the present case with strong 1D anisotropy, weakly incommensurate modulation of the wave vector can be induced by DM interactions due to the lack of inversion symmetry along the chain \cite{Nawa2020, Zvyagin2022}. For the $D$-vector ${\bf D}$ parallel to the $c$ axis, the DM interaction forces the magnetic moments to align perpendicular to {\bf D}. Therefore, we conclude that the moment direction should be close to the $b$ axis. 

\subsection{Critical slowing-down of spin and atomic fluctuations}

\begin{figure}
\includegraphics[width=8.5cm]{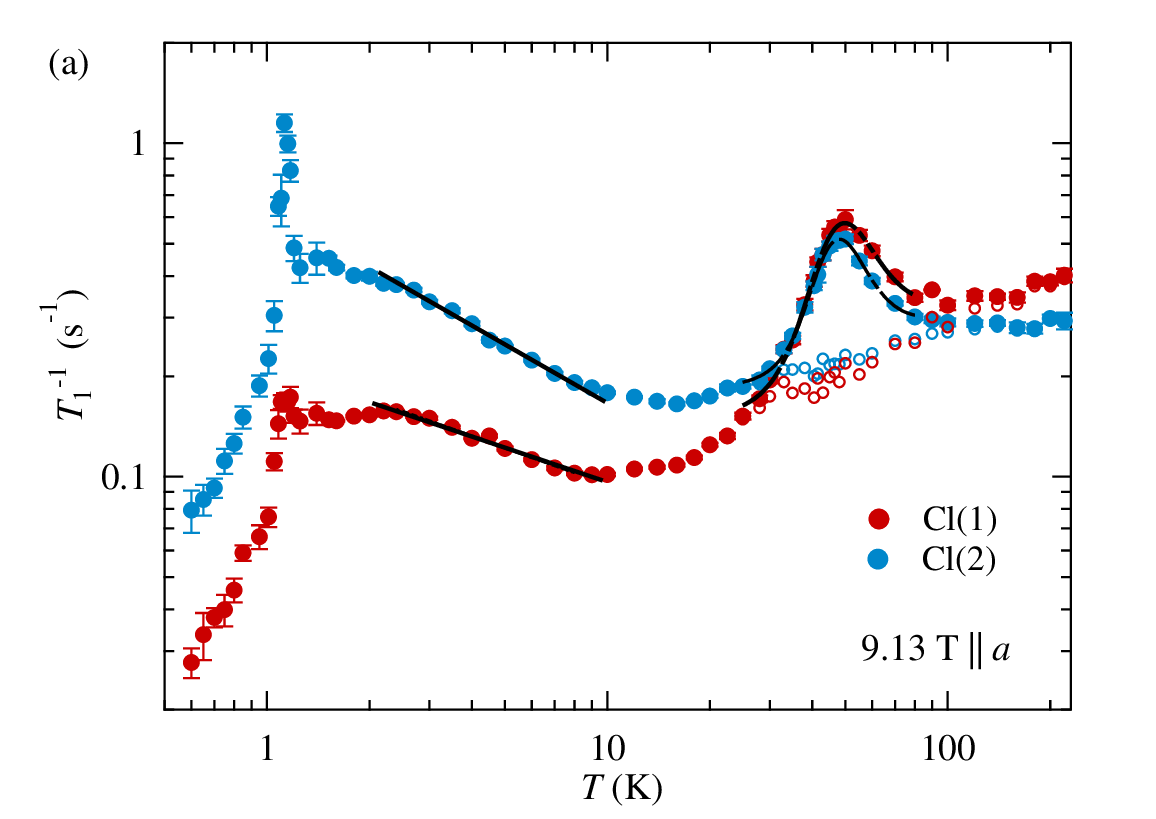}
\includegraphics[width=8.5cm]{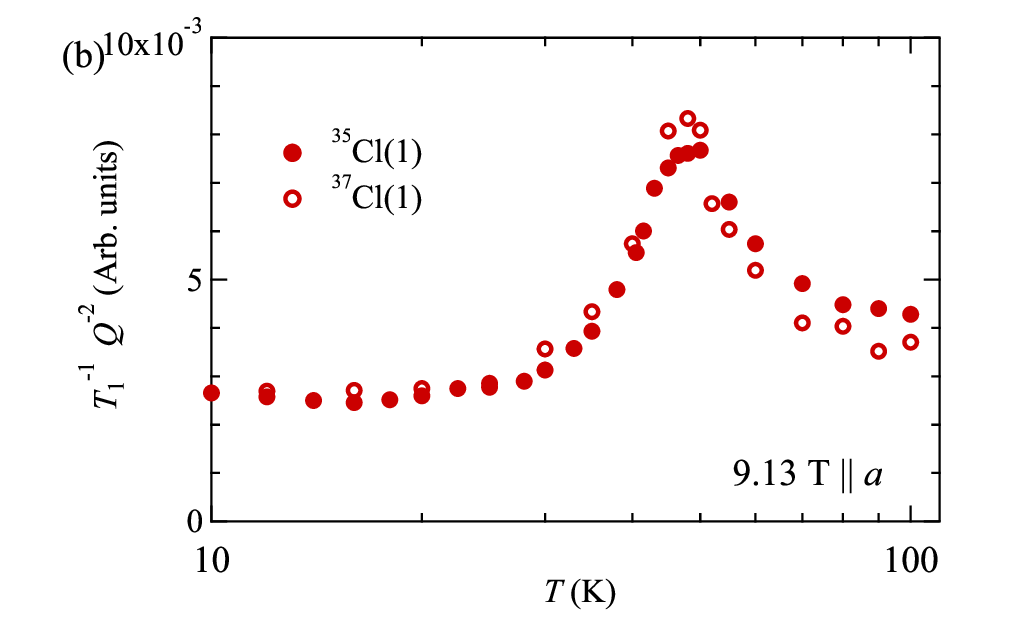}
\caption{(a) Temperature dependence of the nuclear spin-lattice relaxation rate $T_1^{-1}$ measured for the $a$ axis (solid circles) for Cl(1) and Cl(2). Curves around 50 K are the BPP model fitting using Eq. (\ref{eq:BPP}). Open circles represent the result after subtracting the BPP contribution. Solid lines represent the power-law fitting $\sim T^{1/2K_\sigma-1}$ using the Luttinger parameter $K_\sigma$. (b) $T_1^{-1}$ divided by the square of the nuclear quadrupole moment $Q$ for ${}^{35}$Cl and ${}^{37}$Cl NMR.} 
\label{fig:T1_9T_a}
\end{figure}

The nuclear spin-lattice relaxation rate $T_1^{-1}$ measures low-energy excitations in the NMR frequency window through magnetic and electric hyperfine interactions. As shown in Fig. \ref{fig:T1_9T_a}(a), $T_1^{-1}$ exhibits sharp and broad peaks around 1.1 and 50 K, respectively, for $H_0$ = 9.13 T along the $a$ axis. The sharp peak manifests the critical slowing-down of spin fluctuations toward long-range magnetic ordering at $T_{\rm N}$ = 1.1 K. $T_1^{-1}$ drops steeply ($\sim T^2$) below $T_{\rm N}$ where the excitation is dominated by gapless magnons. 

The broad $T_1^{-1}$ peak around 50 K can be attributed to structural fluctuations such as atomic motions instead of phase transition, since there is no signature of symmetry breaking in the NMR spectrum (Fig. \ref{fig:He3}) and specific heat \cite{Hirai2018}. For confirmation, we measured $T_1^{-1}$ of two isotopes $^{35}$Cl and $^{37}$Cl having distinct nuclear quadrupole moments $Q = -8.2$ and $-6.5 \times 10^{-26} {\rm cm}^2$, respectively. As shown in Fig. \ref{fig:T1_9T_a}(b), $T_1^{-1}$ approximately scales to $Q^2$ around 50 K, consistent with the predominant EFG fluctuations \cite{Abragam}. 

\begin{figure}
\centering
\includegraphics[width=8.5cm]{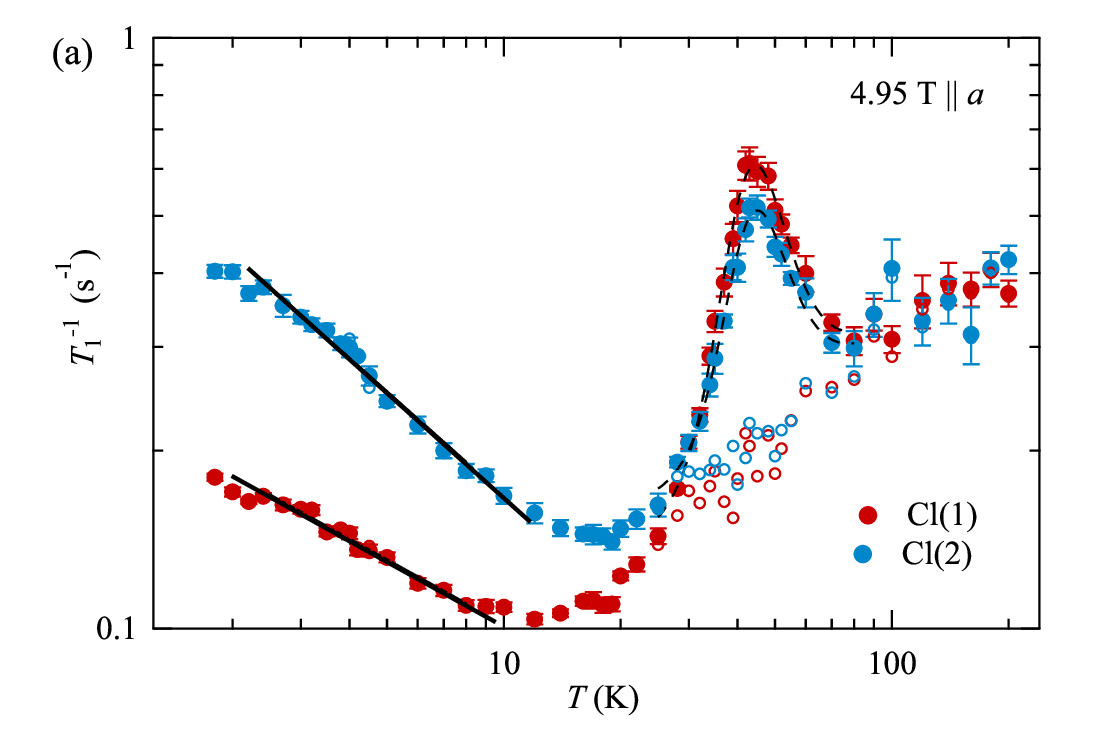}
\includegraphics[width=8.5cm]{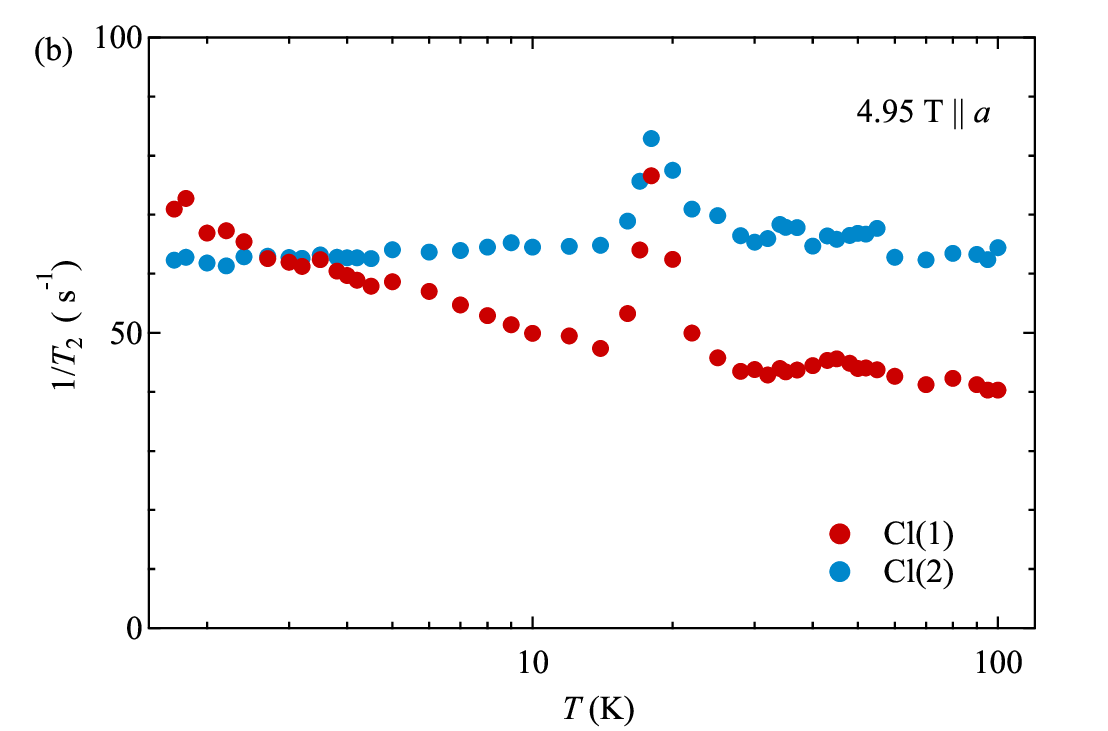}
\hfill
\caption{Temperature dependence of (a) $T_1^{-1}$ and (b) $T_2^{-1}$ at $H_0$ = 4.95 T along the $a$ axis. Broken curves denote the fitting result by the BPP model using Eq.(\ref{eq:BPP}). Solid lines represent the power law $T_1^{-1} \propto T^{1/2K_\sigma-1}$ fitting with the Luttinger parameter $K_\sigma$. }
\label{fig:BPP}
\end{figure}

As the resonance frequency decreases from 38 MHz at 9.13 T to 20 MHz at 4.95 T, the $T_1^{-1}$ peak shifts to 43 K, as shown in Fig. \ref{fig:BPP}(a). Further low frequency ($\sim$ kHz) fluctuations can be observed through $T_2^{-1}$, where the spin-echo intensity decays exponentially with the pulse interval time $\tau = 100 - 1000$ $\mu$s. As seen from the sharp resonance line, $T_2$ is extremely long, despite a paramagnetic Mott insulator, and reaches $\sim 20$ ms for Cl(1). As shown in Fig. \ref{fig:BPP}(b), $T_2^{-1}$ is independent of $T$ above 30 K and exhibits a sharp peak at $T^*$ = 18 K, indicating the freezing of atomic motions. Upon cooling, $T_2^{-1}$ remains constant for Cl(2), while it gradually increases for Cl(1) toward the magnetic order. 

From a structural point of view, the x-ray diffraction measurement on CROC shows a large thermal structure factor of Cl sites at room temperature \cite{Hirai2017}. Although the structural data of CROC are absent at low temperatures, the metastable atomic positions may induce a persistent structural instability due to the large open space. Interestingly, the structural freezing coincides with the onset of antiferromagnetic correlation, as manifested in the $\chi$ peak at 20 K. 

The system crossovers from a quantum TLL regime ($T \ll J$) into a classical diffusive regime ($T \gg J$) with increasing temperature. In an intermediate temperature range of $T \sim J$, spin dynamics of 1D chains is expected to obey the superdiffusive Kardar-Parisi-Zhang (KPZ) universality with characteristic spin correlation \cite{Nardis2019,Dupont2021}. To extract the spin dynamics of the crossover regime, we subtract the structural contribution from $T_1^{-1}$ by assuming the Lorentzian correlation function in the Bloembergen-Purcell-Pound (BPP) model \cite{BPP1947}: 
\begin{equation}
 T_1^{-1} \propto \Bigl[\frac{\tau_c}{1+(\omega\tau_{c})^2}+\frac{4\tau_{c}}{1+(4\omega\tau_{c})^2}\Bigr],
 \label{eq:BPP}
\end{equation}
where $\tau_c$ is the correlation time, $\tau_c = \tau_{0}\exp{(\frac{E_a}{k_{\rm B}T})}$, with the activation energy $E_a$ and the constant $\tau_0$. The experimental result is well fitted by Eq. (2) at the measured frequency (38 and 20 MHz at 9.13 and 4.95 T, respectively), as shown in Figs. \ref{fig:T1_9T_a} and \ref{fig:BPP}, which yields $E_a/k_{\rm B} \approx$ 300 K. The fitting extracts a power law $\sim T^n$ term with the exponent $n = 0.5$ in the temperature range above 20 K. It differs from those of the local spin fluctuations ($n = 0$) \cite{Moriya1956} and the KPZ universality ($n = 2$) \cite{Nardis2019,Dupont2021}. The exponent is rather close to that expected in the spin-drag relaxation of the spin-diffusion regime \cite{Polini2007}. 

\subsection{Spin dynamics in the critical TLL regime}

At low temperatures ($T \ll J$), $T_{1}^{-1}$ would be dominated by spin fluctuations along the chain, where structural fluctuations are exponentially suppressed. According to the fluctuation-dissipation theorem, the imaginary part of dynamical spin susceptibility $\chi({\bf q})$ relates to the dynamical spin structure factor $S^\pm({\bf q})$. $T_{1}^{-1}$ is expressed as \cite{Moriya1956, Berthier2017,Nawa2013, Smerald} 
\begin{eqnarray}
T_{1z}^{-1} &=& 
 \frac{k_{B}T\gamma_{n}^{2}}{2\mu_{B}^{2}\omega}\sum_{\mathbf{q}}[(F_{xx}(\mathbf{q})+F_{yy}(\mathbf{q})\\ \notag
 &+& 2F_{xy}(\mathbf{q}))\chi^{\prime\prime}_\perp(\mathbf{q}) + (F_{xz}({\bf q})+F_{yz}({\bf q}))\chi^{\prime\prime}_{\parallel}({\bf q})]
\label{eq:T1_chi1D}
\end{eqnarray}
where $F_{\alpha\beta}({\bf q})$ ($\alpha, \beta = x, y, z$) is the form factor at the wave vector {\bf q} and defined by $F_{\alpha\beta}({\bf q}) = A_{\alpha\beta}({\bf q})A_{\alpha\beta}(-{\bf q})$ using the Fourier transformed hyperfine coupling $A_{\alpha\beta}({\bf q}) = A_{\alpha\beta}({\bf r}){\rm e}^{i{\bf q\cdot r}}$. $\chi^{\prime\prime}_\perp$ and $\chi^{\prime\prime}_\parallel$ are the imaginary part of the dynamical spin susceptibility perpendicular and parallel to the magnetic field. Here, the $z$ axis is taken along the applied magnetic field direction. In 1D Mott insulators the low-energy $\chi({\bf q})$ is dominated by specific wave vectors close to ${\bf Q_0}$ = (0, $\pi$, 0) \cite{Giamarchi, Chitra1997, Sachdev1994}. A shift from ${\bf Q_0}$ by the magnetic field would be negligible when the Zeeman energy is much lower than the exchange interaction $J$. Instead, the DM interaction induces a small splitting of the spinon dispersion by $\Delta q = 0.07\pi$ \cite{Nawa2020}, which is also omitted in the following analysis for simplicity. 

In a TLL regime, $\chi({\bf q})$ obeys a power law in an anisotropic manner, $\chi_\perp \propto T^{1/2K_\sigma-2}$ and $\chi_\parallel \propto T^{2K_\sigma-2}$ \cite{Chitra1997,Sato2009}. For an isotropic case, both transverse and longitudinal components of $T_1^{-1}$ are independent of temperature ($K_\sigma = 0.5$), as expected for the antiferromagnetic Heisenberg chain \cite{Sachdev1994, Chitra1997}. They become anisotropic as the Ising anisotropy $\Delta$ of the $XXZ$ model deviates from unity. For $K_\sigma > 0.5$, the transverse component is enhanced at low temperatures, whereas the parallel component vanishes and thus becomes negligible. Furthermore, as seen from Eq. (3), $\chi_{\parallel}$ contributes to $T_1^{-1}$ only through the off-diagonal hyperfine coupling. Therefore, $T_1^{-1}$ would be dominated by $\chi_\perp(\mathbf{q})$, as shown in Eq.(1). 

Below 10 K, $T_1^{-1}$ increases with a power law $\sim T^{n}$, as shown in Fig. \ref{fig:T1_9T_a}. The fitting yields $n = -0.36$ for Cl(1) and $-0.59$ for Cl(2) in a magnetic field of 9.13 T along the $a$ axis. Applying the relation $n = 1/(2K_\sigma)-1$ in Eq. (1), we obtained $K_\sigma$ = 0.79 and 1.21 for Cl(1) and Cl(2), respectively. Since the dynamical spin susceptibility is carried by Re spins, $K_\sigma$ should be independent of the probe nuclear spins. Therefore, the observed Cl site dependence comes from the form factor dependence that filters antiferromagnetic fluctuations at the specific wave vector, as typically observed in tetragonal cuprate and pnictide superconductors \cite{Takigawa1991, Smerald}. 

Referring to the crystal structure of CROC in Figs. \ref{fig:crystal} and \ref{fig:hyperfine}, Cl(1) is located between neighboring Re atoms, while Cl(2) is on the Re atom along the $a$ axis. Here we consider antiferromagnetic spin fluctuations at the wave vector $\mathbf{Q_0}$ = (0, $\pi$, 0) along the chain. The staggered hyperfine fields from the second-neighbor Re sites are canceled out at Cl(1), as shown by the site-symmetry analysis in Appendix \ref{Form factor}. Then, the antiferromagnetic correlation at ${\bf q}$ = ${\bf Q_0}$ is filtered through the form factor at Cl(1), while it remains at Cl(2). Therefore, $T_1^{-1}$ measured at Cl(1) is strongly suppressed in contrast to that of Cl(2), despite the uniform hyperfine coupling of Cl(1) greater than that of Cl(2). $T_1^{-1}$ is indeed reversed between Cl(1) and Cl(2) at high temperatures above 50 K, where the uniform ${\bf q} = 0$ mode becomes dominant. 

In the TLL regime, the external magnetic field works as chemical potential and enhances magnetization or $K_\sigma$, as observed in the spin-1/2 ladder system with the spin gap \cite{Horvatic2020}. We have investigated the field dependence of $K_\sigma$ for the Cl sites at $H_0$ = 4.95 T, as shown in Fig. \ref{fig:BPP}. There is no significant field effect on $K_\sigma$, consistent with the negligible effect of the Zeeman energy in the gapless TLL, as listed in Table \ref{table:K_sigma}. Instead, the energy scale of the DM interaction is comparable to the magnetic field and thus impacts on the anisotropy. 

\begin{figure}[t]
\centering
\includegraphics[width=8.5cm]{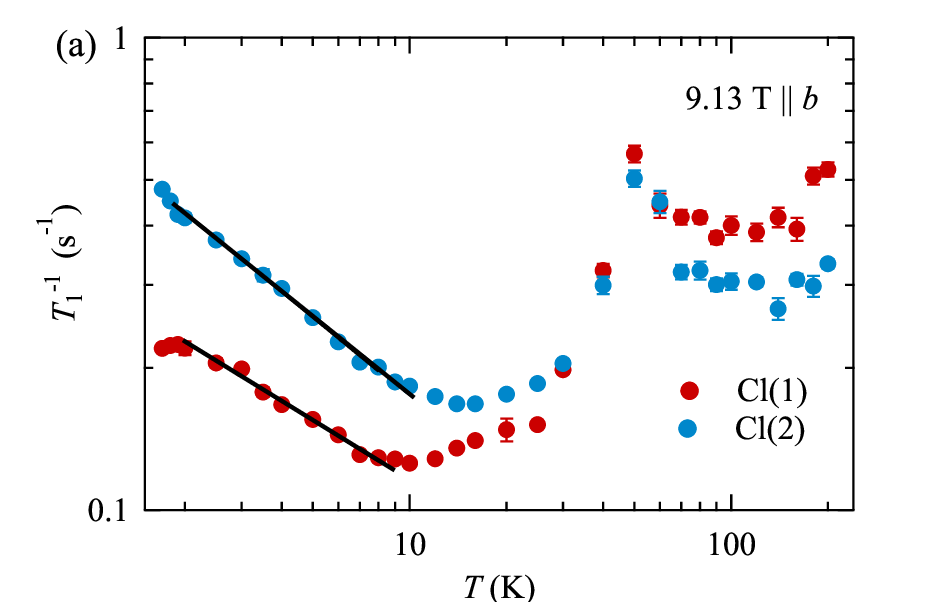}
\includegraphics[width=8.5cm]{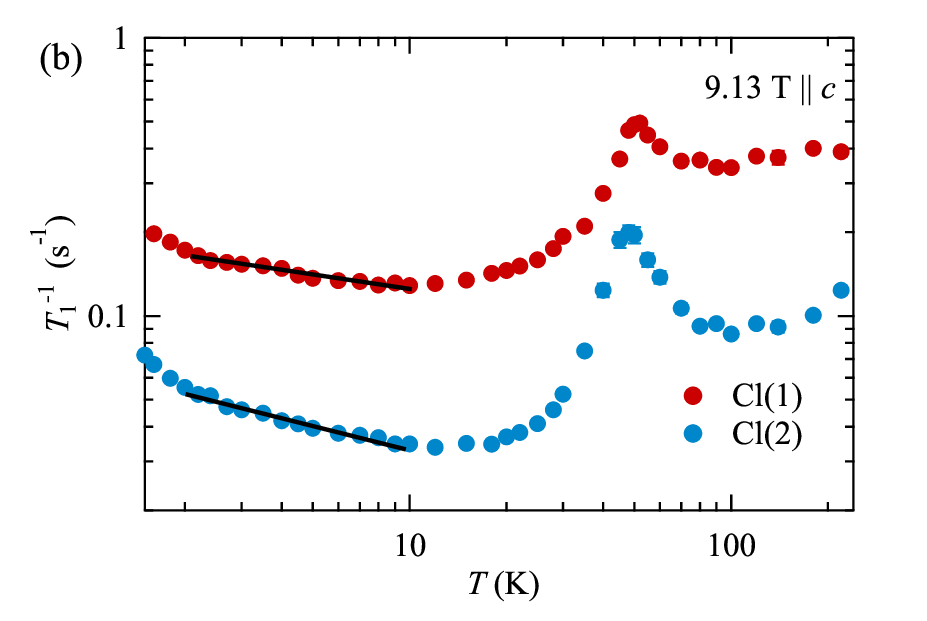}
\hfill
\caption{Temperature dependence of $T_1^{-1}$ for $H_0$ = 9.13 T along (a) $b$ and (b) $c$ axes. Solid lines show the power law $T^{1/2K-1}$ fit with a parameter $K_\sigma$.}
\label{fig:T1_9T_bc}
\end{figure}

\begin{table}[t]
\caption{TLL parameter $K_\sigma$ for Cl(1) and Cl(2) under the different magnetic field orientation and strength. The values in the parentheses are obtained after the RPA analysis \cite{Dupont2018, Horvatic2020}.} 
\label{table:K_sigma} 
\begin{ruledtabular}
\begin{tabular}{ccc}
 & Cl(1) & Cl(2) \\ \hline
$H\parallel a$, 9 T   & 0.79 (0.7) & 1.21 (1.0) \\ 
$H\parallel b$, 9 T   & 0.87 (0.7) & 1.13 (1.0) \\ 
$H\parallel c$, 9 T   & 0.61 (0.5) & 0.70 (0.6) \\
$H\parallel a$, 5 T   & 0.79  & 1.20  \\ 
\end{tabular}
\end{ruledtabular}
\end{table}

\subsection{Anisotropy of $T_1^{-1}$}
Finally, we investigate the anisotropy of $T_{1}^{-1}$ by applying the magnetic field $H_0$ = 9.13 T along the $b$ and $c$ axes, as shown in Fig. \ref{fig:T1_9T_bc}. The result along the $b$ axis is similar to that for the $a$ axis in Figs. \ref{fig:T1_9T_a}(a) and \ref{fig:BPP}, where $T_1^{-1}$ is dominated by atomic fluctuations at high temperatures and then by electric spin fluctuations below 20 K. At low temperatures, we obtain $K_\sigma = 0.87$ and 1.13 for Cl(1) and Cl(2), respectively, as listed in Table \ref{table:K_sigma}. In contrast, the anisotropy of $T_1^{-1}$ for the $c$ axis is not reversed at low temperatures, suggesting that the ${\bf q} = {\bf Q_0}$ mode is less dominant in $T_1^{-1}$ for Cl(1) and Cl(2). Interestingly, $T_1^{-1}$ remains anisotropic in the intermediate temperature range where $T_1^{-1}$ exhibits the broad peak. It may indicate significant spin-lattice coupling through the DM interaction. In the TLL regime below 20 K, we obtained $K_\sigma$, 0.61 for Cl(1) and 0.70 for Cl(2), which is much lower than those in the other directions. The reason can be attributed to the anisotropy of dynamical spin susceptibility, as discussed below. 

\section{Discussion}
The enhancement of $K_{\sigma}$ might be interpreted as the repulsive interaction of spinless fermions based on the $XXZ$ model. In the present system showing isotropic spin susceptibility, the effective spin Hamiltonian would be close to the Heisenberg model with $\Delta \sim 1$. Therefore, the large $K_{\sigma}$ for Cl(2) will originate in the antiferromagnetic correlation toward $T_{\rm N}$, which is not taken into account within the simple TLL theory. 

In the critical regime, the transverse component of $T_1^{-1}$ is expected to follow the power law against reduced temperature, $\sim (T-T_{\rm N})^{-\nu(z_t-D-\eta_t+2)}$, within the dynamical scaling hypothesis \cite{Hohenberg1977, Dupont2018}, where $\nu$ is the correlation length exponent, $z_t$ the dynamical exponent, $D$ the dimensionality, and $\eta_t$ the anomalous exponent. The exponents depend on the universality: e.g. a mean field and a 2D $XY$ model yield $n = -0.5$ and $-0.67$, respectively. Approaching $T_{\rm N}$, the system exhibits a crossover from 1D to 3D regime, leading to the nominal enhancement of $K_\sigma$. 

The effect of critical spin fluctuations on $T_1^{-1}$ can be calculated from the random phase approximation (RPA) below 10 K \cite{Dupont2018, Horvatic2020}. A fitting into our experimental result reduces $K_\sigma$, as listed in Table \ref{table:K_sigma}. The value is still close to unity for Cl(2). The result implies that critical spin fluctuations can be reproduced beyond the mean-field approximation near $T_{\rm N}$. 

The anisotropy of $T_1^{-1}$ can be explained in terms of spinon dynamics under the DM interaction ${\bf D}\cdot{\bf S_i}\times {\bf S_{i+1}}$. In CROC, the DM vector ${\bf D}$ is directed along the $c$ axis within the mirror plane. Spinon dispersion splits by $\pi D/2$ = 310 GHz at $q = 0$ or $\pi$ along the chain, as observed by the EPR measurement \cite{Zvyagin2022}. The excitation branches split with increasing the magnetic field strength along the $c$ axis, whereas they monotonically increase with magnetic field along the $a$ and $b$ axes. The DM interaction induces the magnetic order with the moments directed within the $ab$ plane. As a result, antiferromagnetic fluctuations are enhanced when the magnetic field is applied to the $ab$ plane, while they are suppressed under the field along the $c$ axis where the effect of DM interaction is minimized. Therefore, the critical spin fluctuations are largely suppressed for the $c$ axis, which extracts the intrinsic $K_{\sigma}$ of the TLL regime. 

The field dependence of $K_{\sigma}$ differs from that of the spin-ladder system with the gapped ground state \cite{Horvatic2020, Bouillot2011, Klanjsek2008}. The magnetic field induces the paramagnetic or long-range order phase where the effect of the DM interaction may not be negligible. However, detailed angular dependence measurements of spin fluctuations are absent. Our determination of the TLL parameter through the $T_1^{-1}$ anisotropy measurement will promote further theoretical studies on the anisotropic hydrodynamics of quasi-1D quantum liquids. 

\section{Conclusion}
We have investigated the magnetic ground state and the anisotropic low-energy excitation through the $^{35}$Cl NMR measurement on the quasi-one-dimensional antiferromagnet Ca$_{3}$ReO$_{5}$Cl$_{2}$ with the DM interaction. We determined the nuclear quadrupole splitting and Knight shift tensors by the angular dependence of the $^{35}$Cl NMR spectrum in agreement with the LDA calculation. The nuclear spin-lattice and spin-spin relaxation rates show slow atomic dynamics in the intermediate temperature range. We observed a power-law devolution of spin correlation, characteristic of one-dimensional quantum liquid. The site and angular dependence of the Luttinger parameter comes from the form-factor filtering of the antiferromagnetic correlation induced by the DM interaction. The incommensurate magnetic order occurs below 1 K with the low-lying magnon excitation. 

\section*{Acknowledgements}
We thank a technical support from T. Jinno and useful discussion with H. Yoshioka, S. Capponi, and N. Shannon. We acknowledge the financial support from Grant-in-aid in scientific research by JSPS (No.19H05824, 22H05256, 23H04025, 24H00954, 20H05150, 22H01178, 22H04462). 

\appendix

\section{Nuclear quadrupole splitting}
\label{quadrupole}
In this section, we show a complete set of the angular dependence of nuclear quadrupole splitting $\delta\nu$, which is compared with the DFT calculation.  

The nuclear spin Hamiltonian is expressed as a sum of the magnetic Zeeman interaction and the electric quadrupole interaction $\mathcal{H}_{Q}$: 
\begin{equation}
\mathcal{H} = \mathcal{H}_{Z}  + \mathcal{H}_{Q},   
\end{equation} 
where the Zeeman term is given by
\begin{equation}
\mathcal{H}_{Z} = \gamma\hbar\mathbf{I} \cdot (\mathbf{H_0} + \bf{H_{\rm hf}}) 
\label{nuclear}
\end{equation} 
The hyperfine field is given by $\bf{H_{hf}} = \mathbf{K} \cdot \mathbf{H_0}$ with the Knight shift tensor $\mathbf{K}$. The nuclear quadrupole interaction is given by
\begin{equation}
\mathcal{H}_{Q} = \sum_{\alpha, \beta}\frac{1}{9}\delta\nu_{\alpha\beta} [\frac{3}{2} (I_\alpha I_\beta + I_\beta I_\alpha) - \delta_{\alpha\beta} I^2 ], 
\end{equation}
where the nuclear quadrupole splitting $\delta\nu_{\alpha\beta}$ is defined by 
\begin{equation}
\delta\nu_{\alpha\beta} = \frac{3eQV_{\alpha\beta}}{2I(2I-1)h}
\end{equation}
with the electric field gradient (EFG) $V_{\alpha\beta}$ and the nuclear quadrupole moment $Q$. Under an intense magnetic field, the $^{35}$Cl ($I=3/2$) NMR spectrum splits into three by the interval frequency $\delta\nu_{\alpha\beta}$. 

The second-order electric quadrupole interaction gives a frequency shift of the central $1/2 \leftrightarrow -1/2$ transition in the order of $\nu_Q^2/\gamma_nH_0$. The satellite lines from the $3/2 \leftrightarrow 1/2$ and $-1/2 \leftrightarrow -3/2$ transitions shift equally by the second-order effect. We obtain $\delta\nu_{\alpha\beta}$ by subtracting the lowest satellite frequency from the highest one. 

\begin{figure*}[t]
\centering
\includegraphics[height=0.46\linewidth]{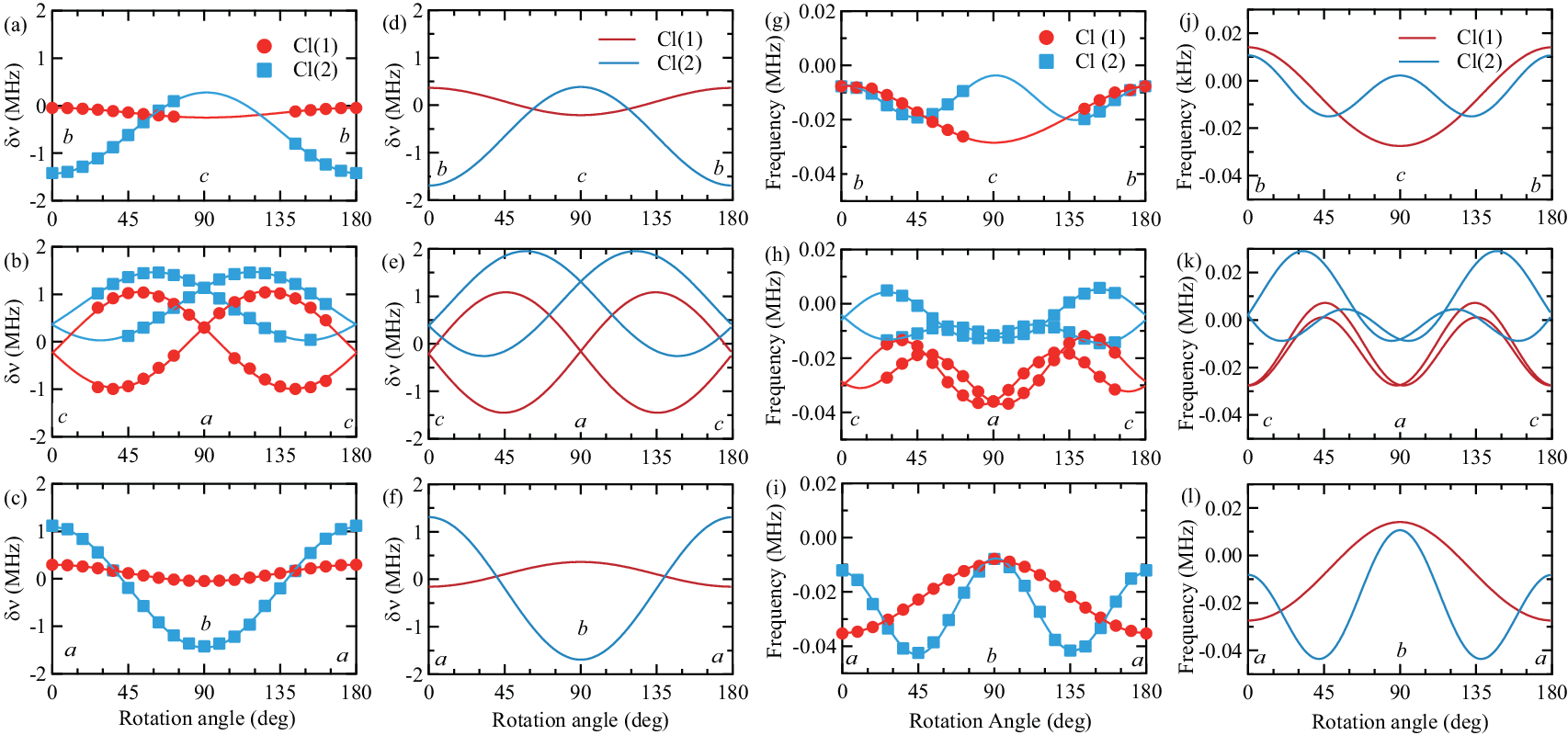}
\label{fig:EFG}
\hfill
\caption{(a-c)Angular dependence of the nuclear quadrupole splitting frequency $\delta \nu$ for two Cl sites at 200 K. The field was rotated along the crystal axis. Solid lines show fitting to Eq. (\ref{eq:efg-fitting}). (d-f) Calculated $\delta \nu$ using LDA. (g-i) Angular dependence of the central frequency for two Cl sites at 200 K. The field was rotated about the crystal axis. Solid curves show fitting to the exact diagonalization calculation of the nuclear spin Hamiltonian including the electric quadrupole interaction and the Zeeman interaction, yielding the Knight shift tensor. (j-l) The second-order quadruple contribution to the central frequency, obtained from the EFG calculation using the LDA functional. }
\end{figure*}

We measured the angular dependence of $\delta\nu_{\alpha\beta}$ about three crystal axes at 20 K and 9.13 T, as shown in Fig. 8. Without detailed analysis below, the profiles of the experimental result in Figs. 8(a,b,c) agree well with those of the LDA calculation in Figs. 8(d,e,f). Thus, we can successfully assign the $^{35}$Cl NMR spectra to two Cl sites with different local environment.

Since both Cl sites are located on the mirror plane normal to the $b$ axis, the $b$ axis must be one of the principal axes of the EFG tensor \cite{V6O13,VO2}. Thus, there is only one set of the spectrum from each Cl site, when we rotate the sample in the crystal plane including the $b$ axis. Then the off-diagonal terms vanish in the $\delta\nu_{\alpha\beta}$ tensor: $\delta\nu_{ab}=\delta\nu_{ba}$ = 0, $\delta\nu_{bc}$ = $\delta\nu_{cb} = 0$. 

The number of NMR lines doubles in the $ca$-plane rotation about the $b$ axis without the spatial inversion symmetry. The rotation about the $b$ axis is analyzed with the Volkoff formula \cite{Volkoff,Slitcher}. On the $ac$-plane rotation with the angle $\theta_b$ measured from the $a$ axis, $\delta \nu_b$ is fitted by 
\begin{equation}
\label{eq:efg-fitting}
\delta\nu_b = \delta \nu_{1,b} + \delta \nu_{2,b} \cos(2\theta_b) + \delta \nu_{3,b} \sin(2\theta_b), \\ 
\end{equation}
where the fitting coefficients are given by
\begin{equation}
\label{eq:efg-fit-coeff}
\begin{split}
\delta \nu_{1,b} = (\delta\nu_{aa} + \delta\nu_{cc})/2 \\
\delta \nu_{2,b} = (\delta\nu_{aa} - \delta\nu_{cc})/2 \\
\delta\nu_{3,b} = -\delta\nu_{ac}.
\end{split}
\end{equation}
Thus, we obtained $\delta\nu_{\alpha\alpha}$ ($\alpha = a, b, c)$ for each crystal axis and the off-diagonal element $\delta\nu_{ac}$, as listed in Table \ref{table:efg}. The diagonalization of the tensor yields the principal components $\delta\nu_{\alpha\alpha}$ ($\alpha$ = X, Y, Z) and the asymmetric factor $\eta= |\delta\nu_{XX} - \delta\nu_{YY}|/\delta\nu_{ZZ}$, where $X$, $Y$, and $Z$ are defined by satisfying $|\delta\nu_{XX}| < \delta\nu_{YY} < |\delta\nu_{ZZ}|$. 
The results are compared with the LDA calculation based on the electronic structure of CROC in Figs. \ref{fig:EFG}(d-f) and Table. \ref{table:efgcal}. Then We can obtain the quadrupole frequency $\nu_Q = \delta\nu_{ZZ}$:
\begin{equation}
    \nu_Q = \frac{3eQV_{ZZ}}{2I(2I-1)h},
\end{equation}
where $V_{ZZ}$ the principal component of the diagonalized electric field gradient tensor \cite{Abragam}. 

\begin{table}[t]
\caption{Nuclear quadrupole splitting frequency $\delta \nu_{\alpha\alpha}$ determined from the angular dependence of the $^{35}$Cl NMR spectrum at 200 K. The tensor for the crystal coordinate ($a, b, c$) is diagonalized with the principal axes ($X, Y, Z$). The asymmetry factor $\eta$ is defined in the text. The $Y$ axis for Cl(1) (or Cl(2)) is directed at $\pm37^\circ$ ($\pm30^\circ$) from the $a$ axis in the $ac$ plane.} 
\label{table:efg}
\begin{ruledtabular}
\begin{tabular}{c c c c c c c c c c c}
  & $\delta\nu_{aa}$ & $\delta\nu_{bb}$ & $\delta\nu_{cc}$ & $\delta\nu_{ac}$ & $\delta\nu_{XX}$ & $\delta\nu_{YY}$ & $\delta\nu_{ZZ}$ & $\eta$ \\ \hline
Cl(1) & 0.29 & $-0.05$ & $-0.24$ & $\pm$0.98 & 0.05 & 0.99 & $-1.04$ & 0.90\\ 
Cl(2) & 1.09 & $-1.45$ & 0.36 & $\pm$0.61 & 0.02 & 1.43 & $-1.45$ & 0.98\\ 
\end{tabular} 
\end{ruledtabular}
\end{table}

\begin{table}[t]
\caption{Nuclear quadrupole splitting $\delta \nu_{\alpha\alpha}$ obtained from the LDA calculation for CROC. The $Y$ (Z) axis for Cl(1) (or Cl(2)) is directed at $\pm40^\circ$ ($\pm33^\circ$) from the $a$ axis in the $ac$ plane.} 
\label{table:efgcal}
\begin{ruledtabular}
\begin{tabular}{c c c c c c c c c c c}
  & $\delta\nu_{aa}^{cal}$ & $\delta\nu_{bb}^{cal}$ & $\delta\nu_{cc}^{cal}$ & $\delta\nu_{ac}^{cal}$ & $\delta\nu_{XX}^{cal}$ & $\delta\nu_{YY}^{cal}$ & $\delta\nu_{ZZ}^{cal}$ & $\eta^{cal}$  \\ \hline
Cl(1) & 0.47 & 0.05 & $-0.52$ & $\pm$1.27 & 0.05 & 1.34 & $-1.39$ & 0.92\\ 
Cl(2) & 1.29 & $-1.71$ & 0.36 & $\pm$1.00 & $-0.22$ & $-1.71$ & 1.93 & 0.77\\
\end{tabular}
\end{ruledtabular}
\end{table}

The angular dependence of the central resonance ($-1/2 \leftrightarrow +1/2$) frequency is shown in Figs. 8(g-i) and compared to the second-order nuclear quadrupole effect obtained from the LDA calculation in Figs. 8(k-l). The good agreement between the experimental result and the calculation allows the site assignment for the two Cl sites. Then we obtained the Knight shift tensor {\bf K} after subtracting the second-order quadrupole contribution by the exact diagonalization of the nuclear spin Hamiltonian of Eq.(\ref{nuclear}).  

We obtained the Knight shift tensor $\mathbf{K}$ arising from the hyperfine interaction with Re electron spins at 200 K: 
\begin{eqnarray}
\mathbf{K} =     
\begin{pmatrix}
 K_{aa} & 0 & K_{ac} \\
 0 & K_{bb} & 0 \\
 K_{ca} & 0 & K_{cc}
\end{pmatrix}
&=& \notag\\
\begin{pmatrix}
 -0.056 & 0 & -0.002 \\
 0 & -0.041 & 0 \\
 -0.002 & 0 & -0.039
\end{pmatrix}
&,&
\begin{pmatrix}
 -0.032 & 0 & -0.001 \\
 0 & -0.033 & 0 \\
 -0.001 & 0 & -0.033
\end{pmatrix}
 \notag
\end{eqnarray}
in \% for Cl(1) and Cl(2), respectively. The diagonalization of the tensor yields $(K_{XX}, K_{YY}, K_{ZZ}) = (-0.039, -0.041, -0.056)\%$ for Cl(1) and $(-0.032, -0.033, -0.033)$ for Cl(2), where $Z$ is directed at $\pm7^\circ$ and $\pm71^\circ$ from the $a$ axis in the $ac$ plane, respectively. We also calculated the magnetic dipole field at Cl sites from Re ions. However, the result disagrees with the experimental result, which indicates that the Knight shift is governed by the transferred hyperfine coupling through the oxygen and calcium ions. Therefore, the hyperfine coupling does not necessarily scale to the atomic distances but depends on the paths through the overlap between the wavefunctions. The following discussion for the form factor is independent of the numerical value of the hyperfine coupling. 

\section{Form factors of dynamical spin susceptibility}
\label{Form factor}
To explain the site dependence of $T_1^{-1}$, we evaluated the form factor of the dynamical spin susceptibility using the symmetry operation in the crystal structure of CROC \cite{Hirai2017}. As shown in Fig. 1, Cl(1) and Cl(2) sites are located between Re ions and on the Re ion, respectively. Here we focus on the original Cl sites with the fractional coordinate of the $Pnma$ $4c$ position, $(x_1, y_1, z_1)$, $y_1 = 1/4$, and the surrounding Re sites ($4c$). Then we semi-quantitatively evaluate the hyperfine coupling based on the local symmetry.

For Cl(1), there are one first neighbor Re site (3.716\AA), Re$^{(1)}_1$, two second ones (4.094\AA), Re$^{(2)}_1$ and Re$^{(2^\prime)}_1$, two third ones (4.892\AA), Re$^{(3)}_1$ and Re$^{(3^\prime)}_1$, and one fourth one (5.868\AA) Re$^{(4)}_1$. 
The fractional coordinates of Re are given by Re$^{(1)}_1$ = $(-x+1, -y+1, 0)$, Re$^{(2)}_1 = (x, y, z)$, Re$^{(2^\prime)}_1 = (x,y-1,z)$, Re$^{(3)}_1 = (x+1/2,y,-z+1/2)$, Re$^{(3^\prime)}_1 = (x+1/2, y-1, -z+1/2)$, and Re$^{(4)}_1 = (-x+3/2, -y+1, z+1/2)$, where $y = 3/4$. 
The relative vectors of the Cl(1)-Re$^{(i)}_1$ bonds are thus written as ${\bf r_1^{(1)}} = (1-x-x_1, 0, -z-z_1)$, ${\bf r_1^{(2)}} = (x-x_1, 1/2, z-z_1)$, ${\bf r_1^{(2^\prime)}} = (x-x_1, -1/2, z-z_1)$, ${\bf r_1^{(3)}} = (x-1/2-x_1, 1/2, -z+1/2-z_1)$, ${\bf r_1^{(3^\prime)}} = x-1/2-x_1, -1/2, -z+1/2-z_1)$, and ${\bf r_1^{(4)}} = (-x-3/2-x_1, 0, z+1/2-z_1)$. 
Here Re$^{(2)}_1$ and Re$^{(2^\prime)}_1$ are connected via mirror reflection normal to the $ac$ plane on Cl(1). 

\begin{figure}
\centering
\includegraphics[width=7cm]{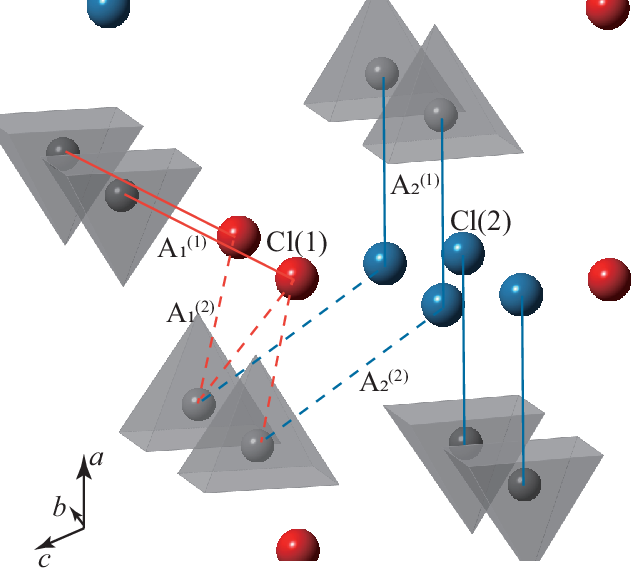}
\hfill
\caption{Geometry of hyperfine interaction between Cl nuclear spins and Re sites in Ca$_{3}$ReO$_{5}$Cl$_{2}$. Solid and dashed lines indicate nearest and next nearest Re-Cl bonds.}
\label{fig:hyperfine}
\end{figure}

Thus the hyperfine coupling tensor can be expressed as
\begin{eqnarray}
{\bf A^{(1)}_1} &=&     
\begin{pmatrix}
 A_{1aa}^{(1)} & 0 & A_{1ac}^{(1)} \\
 0 & A_{1bb}^{(1)} & 0 \\
 A_{1ca}^{(1)} & 0 & A_{1cc}^{(1)}
\end{pmatrix}
\\
{\bf A^{(2)}_1} &=&     
\begin{pmatrix}
 A_{1aa}^{(2)} & A_{1ab}^{(2)} & A_{1ac}^{(2)} \\
 A_{1ba}^{(2)} & A_{1bb}^{(2)} & A_{1bc}^{(2)} \\
 A_{1ca}^{(2)} & A_{1cb}^{(2)} & A_{1cc}^{(2)}
\notag
\end{pmatrix}
\\
{\bf A^{(2^\prime)}_1} &=&     
\begin{pmatrix}
 A_{1aa}^{(2)} & -A_{1ab}^{(2)} & A_{1ac}^{(2)} \\
 -A_{1ba}^{(2)} & A_{1bb}^{(2)} & -A_{1bc}^{(2)} \\
 A_{1ca}^{(2)} & -A_{1cb}^{(2)} & A_{1cc}^{(2)}
\end{pmatrix}
.
\end{eqnarray}
${\bf A^{(3)}}$, ${\bf A^{(3^\prime)}}$, and ${\bf A^{(4)}}$ are written similar to ${\bf A^{(2)}}$, ${\bf A^{(2^\prime)}}$, and ${\bf A^{(1)}}$, respectively. 
The net hyperfine coupling tensor is expressed as a sum of the hyperfine paths, ${\bf A_1} = \sum_i {\bf A_1^{(i)}}$. 
Taking Fourier transformation 
\begin{equation}
{\bf A_1(q)} = \sum_i {\bf A_1^{(i)}}({\bf r_j}){\rm e}^{i{\bf q \cdot r_j}}, 
\end{equation} 
we obtain the uniform mode of the hyperfine coupling tensor for Cl(1)
\begin{eqnarray}
{\bf A_1(0)} \simeq     
\begin{pmatrix}
 A_{1aa}^{(1)}+2A_{1aa}^{(2)} & 0 & A_{1ac}^{(1)}+2A_{1ac}^{(2)} \\
 0 & A_{1bb}^{(1)}+2A_{1bb}^{(2)} & 0 \\
 A_{1ca}^{(1)}+2A_{1ca}^{(2)} & 0 & A_{1cc}^{(1)}+2A_{1cc}^{(2)}
\end{pmatrix}
\notag
\end{eqnarray}
where we omit the higher order terms for simplicity. The staggered mode along the $b$ axis with the specific wave vector ${\bf Q_0} = (0, \pi, 0)$ is expected to dominate the low-energy spin correlation in the 1D chain. The hyperfine tensor at ${\bf Q_0}$ is calculated as 
\begin{eqnarray}
{\bf A_1(Q_0)} \simeq     
\begin{pmatrix}
 A_{1aa}^{(1)} & 2iA_{1ab}^{(2)} & A_{1ac}^{(1)} \\
 2iA_{1ba}^{(2)} & A_{1bb}^{(1)} & 2iA_{1bc}^{(2)} \\
 A_{1ca}^{(1)} & 2iA_{1bc}^{(2)} & A_{1cc}^{(1)}
\end{pmatrix}
\notag
.
\end{eqnarray}
The second neighbor interactions remain for ${\bf A_1(0)}$ but vanished for ${\bf A_1(Q_0)}$ in the diagonal components. Namely, antiferromagnetic spin fluctuations are partly filtered through the hyperfine form factor at Cl(1). 

As for Cl(2) = $(x_2, y_2, z_2)$, $y_2=3/4$, there are one nearest neighbor (3.555\AA) and one second neighbor (4.350\AA) Re site. We omit the higher order terms with the distance exceeding 5 \AA. Similar to Cl(1), the Cl(2)-Re vectors are expressed as ${\bf r_2^{(1)}} = (x-1/2-x_2, 0, -z+1/2-z_2)$ and ${\bf r_2^{(2)}} = (x-x_2, 0, z-z_2)$. Thus both the uniform and staggered mode of the hyperfine coupling tensor ${\bf A(q)}$ (${\bf q} = {\bf 0}, {\bf Q_0}$) is expressed as 
\begin{eqnarray}
{\bf A_2(q)} \simeq
\begin{pmatrix}
 A_{2aa}^{(1)}+A_{2aa}^{(2)} & 0 & A_{2ac}^{(1)}+A_{2ac}^{(2)} \\
 0 & A_{2bb}^{(1)}+A_{2bb}^{(2)} & 0 \\
 A_{2ca}^{(1)}+A_{2ca}^{(2)} & 0 & A_{2cc}^{(1)}+A_{2cc}^{(2)}
\end{pmatrix}
\notag
.
\end{eqnarray}
The hyperfine form factor along the field direction, $F_{\parallel}$ is calculated as $F_{\parallel} = \sum_\alpha {A_{\alpha \beta}(q)A_{\alpha\beta}(-q)}$. 
For $r_1^{(1)} > r_2^{(1)}$, $A_{2\alpha\beta}^{(1)}$ is greater than $A_{1\alpha\beta}^{(1)}$. Since the staggered spin fluctuations at ${\bf Q_0}$ are dominant at low temperatures, the reversed $T_1^{-1}$ values of Cl(1) and Cl(2) can be attributed to the partial cancellation of the form factor at Cl(1) due to the antisymmetric spin fluctuations at the second neighbor Re spins. 

\bibliography{bib}

\end{document}